\documentclass[namedreferences]{SSRv}
\usepackage{epsfig}
\usepackage{graphicx}
\usepackage{times}
\usepackage{makeidx}

\def\sci{{Science}}

\def\grl{{Geo\-phys.\ Res.\ Lett.}}

\def\jgr{{J.\ Geo\-phys. Res.}}
\def\nat{{Nature}}
\def\prl{{Phys.\ Rev.\ Lett.}}

\def\ssr{{Space Sci.\,Rev.}}
\def\rmp{{Rev.\ Mod.\ Phys.}}

\def\pop{{Phys.\ Plasmas}}

\newcommand{\w}[1]{\mbox{\boldmath{$#1$}}}

\makeindex

\begin{document}

\begin{article}
\begin{opening}

\title{PHYSICS OF ELECTRIC DISCHARGES  IN ATMOSPHERIC GASES: 
AN INFORMAL INTRODUCTION }

\author{RUDOLF A.  \surname{TREUMANN}$^{1,2}$,
	ZBIGNIEW  \surname{K{\L}OS}$^{3}$,
      MICHEL    \surname{PARROT}$^{4}$
        }

\runningauthor{TREUMANN, K{\L}OS AND PARROT}
\runningtitle{Atmospheric discharge physics}

\institute{$^{1}$ Department of Geophysics and Environmental Sciences, Munich University, Munich, Germany \email{treumann@geophysik.uni-muenchen.de, treumann@issibern.ch}\\
           $^{2}$ Department of Physics and Astronomy, Dartmouth College, Hanover NH 03755, USA\\
           $^{3}$ Space Research Centre, Polish Academy of Sciences, Warsaw, Poland
                  \email{klos@cbk.waw.pl}\\
           $^{4}$ Laboratoire de Physique et Chimie de  l'Environnement-CNRS, F-45071 Orleans, France \email{mparrot@cnrs-orleans.fr}
                           }

\date{Received ; accepted }

\begin{abstract}

A short account of the physics of electrical discharges in gases is given from the viewpoint of its historical evolution and application to planetary atmospheres. As such it serves as an introduction to the papers on particular aspects of electric discharges contained in this issue, in particular in the chapters on lightning and the discharges which in the last two decades have been observed to take place in Earth's upper atmosphere. In addition to briefly reviewing the early history of gas discharge physics we discuss the main parameters affecting atmospheric discharges like collision frequency, mean free path and critical electric field strength. Any discharge current in the atmosphere is clearly carried only by electrons. Above the lower boundary of the mesosphere the electrons must be considered magnetized with the conductivity becoming a tensor. Moreover, the collisional mean free path in the upper atmosphere becomes relatively large which lowers the critical electric field there and more easily enables discharges than at lower altitudes. Finally we briefly mention the importance of such discharges as sources for wave emission.

\end{abstract}

\end{opening}
 \vspace{-0,5cm}
{\small

\section{A brief historical overview}
\label{History} 

\noindent The physics of electrical discharges in gases has a long history \cite[1998]{Bowers1991} reaching back into the 17th century;  in 1672 Otto von Guericke\index{von Guericke, Otto} reported the production of static electric sparks when rubbing sulphur balls with the hand. Three years later the astronomer Jean Picard\index{Picard, Jean} reported that shaking a mercury barometer caused the tube to glow. At the end of that century (around 1698) sparks had been extracted from rubbed amber and formed the subject of intense investigation by the Comte Charles Fran{\c c}ois de Cisternay du Fay.\index{du Fay, Comte Charles}  Based on Picard's observation Johann Heinrich Winckler\index{Winckler, Johann Heinrich} in Leipzig invented what we can call a `fluorescent tube' in 1745, and in 1802 Vasilii Petrov\index{Petrov, Vasilii} in St Petersburgh produced electric arcs between two carbon electrodes independently of, but almost at the same time as, Sir Humphry Davy \cite{Knight1992} in London. Davy\index{Davy, Sir Humphry} invented the carbon arc lamp, presenting it in 1807 to the Royal Institution, and established the subject of arc physics \cite{Davy1807}. For their experiments they both used the chemical battery invented in 1799 by Count Alessandro Volta. \index{Volta, Count Alessandro}

The famous experiments of Benjamin Franklin in 1750\index{Franklin, Benjamin} on sparks led him to coin the notions of positive and negative charging which  survive until today and to invent the lightning rod. Mikhail V. Lomonossov\index{Lomonossov, Mikhail V.} in 1743 \cite{Lomo1748} had already suggested that lightning, and the polar lights (aurorae)\index{ionosphere!aurora} as well, would be atmospheric electric discharge phenomena. With respect to the similarity between lightning and sparks they were both right, while they could not know that auroral discharges are phenomena quite different from spark discharges. Charles-Augustin de Coulomb (1785)\index{Coulomb, Charles-Augustin de} discovered dark discharges \cite{Gillmor1971}. In the 19th century interest in gas discharges increased due to Michael Faraday's\index{Faraday, Michael} and James Clerk Maxwell's\index{Maxwell, James Clerk} formulation of the physics of electromagnetic phenomena. Faraday himself even experimented with sparks and glow (or coronal) discharges, thereby uncovering the dependence of the latter on the pressure in the tube \cite[1834]{Faraday1833}. 
It is interesting to note that such investigations were made possible by improved glass blowing techniques the evacuation of glass tubes and the separation of gases. 

Experimenting on dilute gases, Johann Hittorff \cite{Hittorff1879}\index{Hittorff, Johann} determined the \index{conductivity!of air} electrical conductivity of air and different gases which, in 1900, culminated in Paul Drude's\index{Drude, Paul} seminal formula for the electrical conductivity\index{gas!conductivity}\index{gas!Drude formula} of a gas, introducing the concept of collision frequency $\nu_c$.  The collision frequency received its physical explanation when Ernest Rutherford\index{collisions!Rutherford cross section} found the now famous Rutherford collisional cross section, $\sigma_c$, in terms of which the collision frequency can be written as $\nu_c= n\sigma_cv$, where $n$ is the gas density, and $v$ the velocity of the moving particle. Equivalently this allowed the introduction of the collisional free flight distance $\lambda_{\rm ff}= v/\nu_c$\index{particles!mean free path} of a particle between two collisions. When $v$ was taken to be the thermal velocity $v_{\rm th}$ this became the mean free path $\lambda_{\rm mfp}= v_{\rm th}/\nu_c=1/n\sigma_c$.  

Hittorff somewhat later published the first investigations on cathode rays which were crucial for the later formulation of atomic physics. Indeed, any deeper understanding of gas discharges had to wait until the advent of atomic physics, which became possible only via reference to the investigation of gas discharges.  William Crookes \cite{Crookes1879}\index{Crookes, William} found that cathode rays were deflected by magnetic fields and called them the fourth state of matter, claiming that their investigation would lead to the deepest insight into the nature of matter. This prediction was in fact true. From the deflection of cathode rays in a magnetic field and based on the electrodynamics of charged particle motions in electromagnetic fields, Sir Joseph Thomson \cite{Thomson1897}\index{Thomson, Sir Joseph} determined the elementary charge-to-mass ratio which led to the discovery of the electron. One year earlier Wilhelm Conrad Roentgen \cite{Roentgen1895} experimenting with cathode rays had discovered X-rays.

The last decade of the 19th century and the first decade of the next were devoted to investigations of spectral phenomena with the help of dilute gas tubes, leading on the one hand to the determination of the Johann Balmer \cite{Balmer1885},\index{Balmer, Johann} Walther Ritz \cite{Ritz1908},\index{Ritz, Walter} Theodore Lyman \cite{Lyman1914}\index{Lyman, Theodore} and  Friedrich Paschen \cite{Paschen1908} \index{Paschen, Friedrich} series of spectral line, and on the other hand to the precise mapping of the temperature dependence of black body radiation, and the energetic  of collisional excitation of electrons in atoms at specific energies by James Franck and Gustav Hertz \cite{Franck1914}. \index{Franck, James}\index{Hertz, Gustav} These investigations culminated in the formulation of quantum and atomic theory, going hand in hand with the development of gas discharge physics. It is thus no surprise that most of the Nobel laureates of the first three decades of the 20th century had worked on the physics of gas discharges, among them Rutherford, John S. Townsend, Owen W. Richardson, Karl Compton, Irving Langmuir and Peter Debye. \index{Compton, Karl} In particular Townsend's \cite[1915]{Townsend1901} \index{Townsend, John S.}and Richardson's \cite[1928]{Richardson1908} investigations of the different types of discharges were crucial for the further development of gas discharge physics. Clearly the peak of gas discharge physics coincided with the development of atomic physics and quantum mechanics, i.e. with the quantum theory of the atom. \index{Richardson, Owen W.}\index{Rutherford, Lord Ernest }  

The first three decades of the 20th century saw the physics of gas discharge phenomena blossoming. With the investigation of the various processes of the ionisation of gases, by either radiation or collisions, the understanding of atomic spectra and of the excitation processes achieved quite a high level of sophistication and maturity.  At the same time the physics of gaseous discharges bifurcated into two quite separate disciplines, proper \index{gas!discharge} gas discharges and \index{plasma} plasma physics. The notion of a plasma was introduced by Irving Langmuir in 1927 \cite{Mott-Smith1971} reserving it to highly \index{gas!ionized} ionised gases,\index{Langmuir, Irving} with only a small residual admixture of a non-ionised neutral component or even lacking such a component. Due to the dominance of electromagnetic interactions in a plasma, the physics of plasmas turned out to be completely different from that of neutral gases. Peter Debye and Erich H{\"u}ckel \cite{Debye1923}\index{Debye, Peter}\index{H\"uckel, Erich} showed that electrical charges in a plasma are screened beyond a distance $\lambda_D=v_e/\omega_{pe}$, termed the \index{Debye length} Debye radius, where $v_e=v_{{\rm th}e}$ is the electron thermal velocity, and $\omega_{pe}$ is the \index{frequency!plasma frequency} electron plasma frequency, the oscillation frequency of electrons around their equilibrium position (discovered by Langmuir 1923). 

The notion that the upper atmosphere of the Earth must be electrically conducting goes back to Carl Friedrich Gauss (1776-1855)\index{Gauss, Carl Friedrich} who, expanding the Earth's magnetic field \index{field!geomagnetic} in spherical harmonics, discovered that it contained a substantial component due to outer sources which he suspected to be currents somewhere at large altitude above the surface. It was only in 1902 that Oliver Heaviside \cite{Nahin1990,Appleton1947,Watson-Watt1950} proposed the existence of a high-altitude conducting layer that reflects \index{waves!radio} radio waves. This was confirmed experimentally in 1924-1927 by Edward V. Appleton \cite{Appleton1947}.\index{Appleton, Edward V.} The formation and nature of this layer was understood when Sydney Chapman \cite{Chapman1931}\index{Chapman, Sidney}  realised that solar UV radiation was capable of ionising the upper atmospheric layers at around 100-130 km \index{ionosphere!Chapman layer} altitude, being absorbed there. H. Kallmann \cite{Kallmann1953} ultimately proved that taking into account the full solar spectrum the ionisation effectively reached down to below 80 km altitude. With the advent of the space age it became clear that atmospheric constituents below 80 km can  also be ionised by precipitating energetic particles from the Sun and the magnetosphere and by cosmic rays. These processes are reviewed elsewhere in this issue (Chapter 2 and Chaper 6). \index{Heaviside, Oliver}

In the early 1930s several extended reviews were written on this subject, among them the famous accounts of Compton and Langmuir \cite[1931]{Compton1930}. After the golden twenties and thirties the physics of proper gas discharges lost its importance in fundamental physics, becoming a branch of applied physics, while plasma physics -- its fully ionized twin -- shifted into centre stage. This was mainly for two reasons, the interest in fusion research and the advent of the space age. Space plasma physics became attractive to both the public and politicians. Still, interest in ionospheric physics and in the mechanism of lightning enabled the physics of gas discharges to stay alive throughout  this period of relative drought. 

In the last twenty years gas discharges became more interesting again, in relation to \index{TLE}natural transient luminous events ({\footnotesize TLE}s) \cite{Franz1990,Sentman1995,Neubert2003} that had been discovered in the Earth's upper atmosphere above thunderstorms. Such spectacular phenomena are \index{sprites} \index{elves} \index{jets} sprites, electric jets, and elves \cite{Fullekrug2006}; evidence for lightning and related electrical effects on other planets, in planetary rings and even in astrophysical objects, accumulated. In addition, the relation of TLEs to space weather phenomena has provided another impetus to their investigation. [The processes of the generation and separation of charges which are required before a discharge can be generated are reviewed in this issue \cite{Roussel2008,Yair2008} and also form the content of Chapters 2 and 6. Some of the theory and the observation of such effects as blue jets and sprites are summarized in Chapter 7 \cite{Mishin2008,Mika2008}.]  Because of the importance of such effects the physics of gaseous discharges is far from being exhausted, in particular in view of the newly discovered fine structuring of streamers \cite{Ebert2006}. 

There is one particular aspect of this that in the future will certainly regain more attention. This is the problem of the energetic coupling between near-Earth space and the atmosphere. More generally, the question of how a planetary atmosphere couples to its space environment poses a problem that is of fundamental importance for climatology and the understanding of how energy is fed into processes that are important for the planetary climate from the outside, the interplanetary and magnetospheric plasmas \index{magnetosphere} through coupling via the ionosphere \index{ionosphere} to the atmosphere. Sprites, elves and jets suggest that this is indeed an important problem which has only just been realised and has not yet been investigated to the extent that it deserves.  There are extended recent reviews of these phenomena available \cite[for instance]{Fullekrug2006,Pasko2007}.

\section{Parameters and Theory}\label{parameters}
\noindent The electric discharge is the final step of a whole sequence of processes going on in a gas that is subject to ionisation and internal gas dynamics. Most of these processes have been or will in subsequent sections in this issue be described for Earth and other planets surrounded by an atmosphere or some equivalent to that. At the end of all these processes there is an electric field of strength ${\w E}$ that is maintained in some way over a sufficiently long time, and positive and negative electrical charge layers that are separated from each other by a distance $L$. This electric field can be maintained only up to a certain strength $|{\w E}|\leq|{\w E}|_{\rm crit}$ which, when exceeded, causes a spectacular breakdown of the field\index{field!critical electric}. The processes of of electric field build-up are related to the various processes of ionisation and fake up considerable space in this issue. Here we restrict ourselves to a very basic and by no means complete discussion of the breakdown processes. Several aspects of these will be given in great detail in Chapter 7. 

The atmosphere is considered to be collisional, with the collision frequency $\nu_c$ being determined mainly by collisions between electrons and the various neutral components of the atmosphere. A more sophisticated theory must take into account the composition of the atmosphere including aerosols, various excitation and ionisation processes such as, for instance, cosmic ray ionisation or frictional ionisation, the presence of a magnetic field, the horizontal and vertical dynamics of the atmosphere, its water content, temperature distribution, and its altitudinal inhomogeneity. All these factors (including those we are not considering here) enter into the real discharge process, and some of them are considered in great detail elsewhere in this issue \cite{Roussel2008}. Here we merely mention them as hard-to-treat complications of a realistic atmospheric discharge process encountered, for instance, in a lightning discharge or the recently discovered high-altitude discharges in sprites, blue jets and giant jets.\index{jets!blue} \index{jets!giant}

In the simplest model of a resistive atmosphere the collisions are treated classically (and we will not deviate from this in this introductory paper). This implies that $\sigma_c\approx 4.5\times 10^{-18}$ m$^{-2}$ \cite{Aguilar1990,Bernshtam2000} is the constant classical two-body \index{collisions!cross section} collisional cross section. In addition the neutral gas density obeys the barometric law $n_n(h) = n_0 \exp [-m_n g(h) h/ $ $k_B T_n(h)]$, where the subscript $n$ means neutrals, $h$ is the height above sea level, $n_0$ the density at ground level, $g(h)$ the gravitational acceleration at altitude $h$, $m$ the mass, $T(h)$ the temperature at $h$, and $k_B$ Boltzmann's constant. By contrast in a fully ionised plasma the cross section is the Rutherford or (when averaged over the thermal distribution function) the Spitzer-Braginskii cross section \cite[Chapter 4]{Baumjohann1996}\index{gas!barometric law} which strongly  depends on plasma density and particle speed, thus implying a completely different physics. Since in the uppermost layers of the atmosphere\index{gas!atmosphere} and in particular in the ionosphere the gas is weakly ionised with an admixture of electrons and ions, the relevant cross section will be a mixture of the classical and Spitzer cross sections. In the neutral atmosphere on the other hand several other effects have to be considered which are related to the excitation cross sections of the different gas molecules, the molecular nature of the gas and the composition of the atmosphere at different altitudes, as well as the different ionisation energies. Moreover, the eventual presence of an electric field and the background convective wind motions of the gas as well as the effects of the external radiation change the conditions existing. All such effects are ignored here but for realistic cases should be taken into account in their relative importance. 
\begin{figure}%
\centerline{\includegraphics[width=9cm]{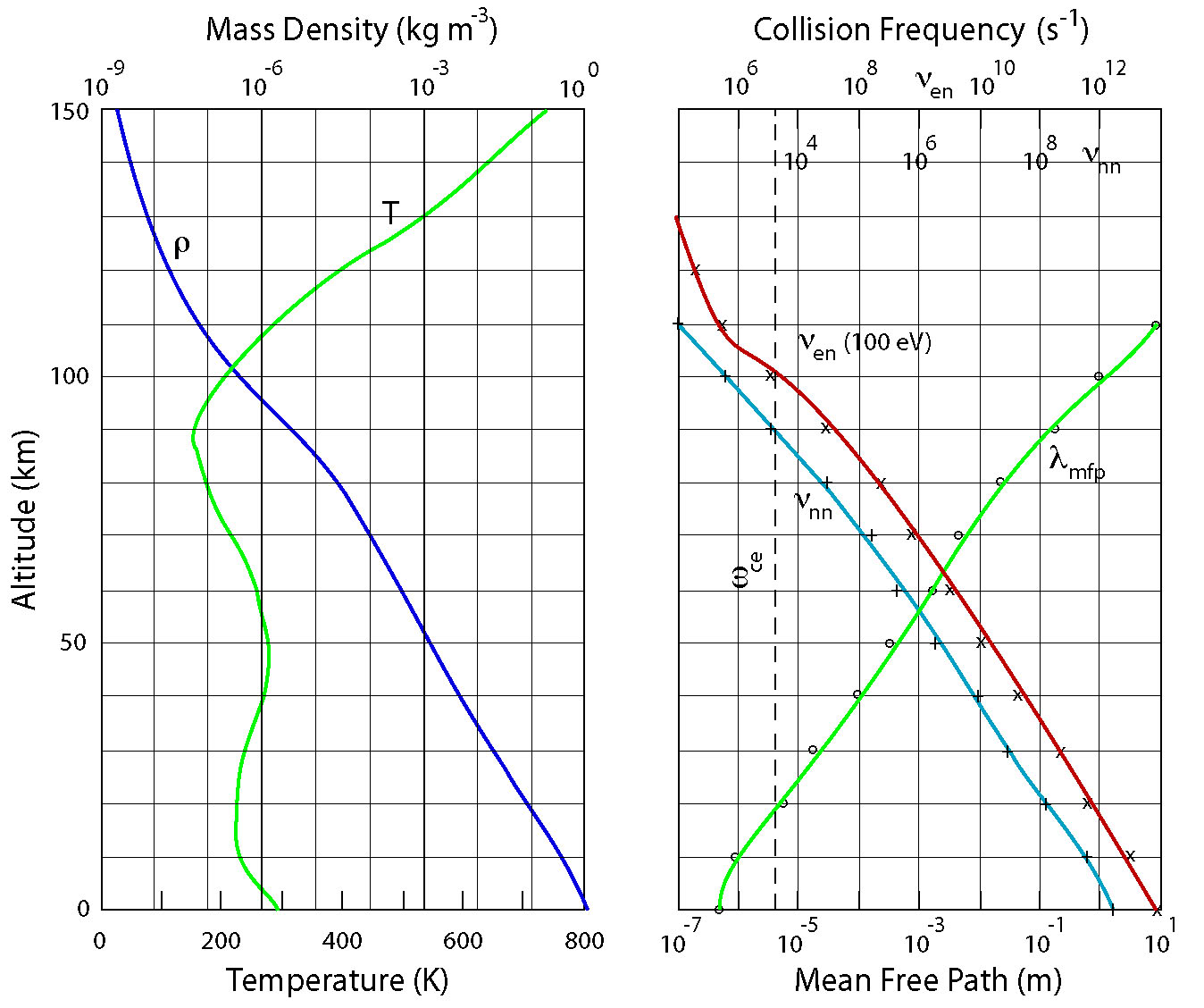}} \caption{{\it Left}: Average altitude profiles  of temperature $T(h)$ and mass density $\rho(h)$ in the Earth's atmosphere consisting of 80\% N$_2$ and 20\% O$_2$. {\it Right}: Rough models of collisional mean free path $\lambda_{\rm mfp}$ based on the data on the left  for neutral-neutral binary collisions with collision frequency $\nu_c=\nu_{nn} (h)$, and the electron-neutral collision frequency for collisions between electrons of 100 eV energy and neutrals as functions of altitude above ground level. The light line shows the altitude dependence of the binary collisional mean free path $\lambda_{\rm mfp}(h)$. In the middle atmosphere the mean free path and collision frequencies vary about exponentially with altitude. The dashed vertical line shows the \index{frequency!cyclotron frequency}  electron cyclotron frequency $\omega_{ce} \simeq 4.5\times 10^6$ Hz which is nearly constant over this height range. The estimate of the electron collision frequency neglects excitation and Coulomb interactions which should be included above roughly 10 eV electron energy for a more realistic model. This would lead to a substantial modification of $\nu_c=\nu_{en}$.}\label{tre-fig-temper}
\end{figure}

With the above notations the collision frequency becomes
\begin{equation}\label{tre-eq-1}
\nu_c (h)= \nu_0 \exp \left[-\frac{m_ng(h)h}{k_BT_n(h)}\right]\sqrt{\frac{T_n(0)}{T_n(h)}}, \qquad \nu_0\equiv n_0\sigma_c\sqrt{\frac{2k_BT_0}{m_n}}.
\end{equation}
On Earth, the temperature and density of air\index{gas!atmosperic temperature}\index{gas!atmosperic density} at ground level are roughly $T_0\approx 300$ K and $n_0\simeq 2\times 10^{25}$ m$^{-3}$. The latter value depends on what is assumed for the composition of air. In the above formula the height dependence of the gravitational acceleration $g(h)\simeq g(0)(1-2h/{\rm R_ E})$ can safely be ignored since in the altitude range from 0 to 100 km the variation of the ratio of height to Earth radius R$_{\rm E}$ is $\lesssim 1\%$. The temperature (see Figure \ref{tre-fig-temper})  in this altitude range changes by a factor of  $T_n(h)/T_0\leq 3$. Thus its change is important only in the exponential factor. At zero level, the Earth's surface, we have for the neutral-neutral collision frequency $\nu_0\sim 10^9$ Hz.

Figure \ref{tre-fig-temper} summarises the vertical profile of the Earth's atmospheric temperature, mass density and binary collision frequency in a simplified way with the neutral-neutral collision frequency $\nu_c=\nu_{nn}$ and electron-neutral collision frequency $\nu_c=\nu_{en}$ calculated for an 80\% nitrogen and 20\% oxygen mass-dominated atmosphere. The reason for the latter being  large is the proportionality of the collision frequency to the thermal velocity $v_e$ of the electrons which are generated by ionisation and therefore have energies $>$\, a few eV. At higher electron energies as those they gain during a discharge this collision frequency becomes obsolete because other processes (excitation, Coulomb interaction)\index{particles!excitation} \index{particles!ionisation}set on. Since the Earth is magnetized the collision frequency is, in principle, a tensor \cite[e.g.]{Rycroft2008} giving rise to a tensorial electrical conductivity ${\w \sigma} = (\sigma_\|, \sigma_P, \sigma_H)$, where ${\sigma_P}$ is the conductivity perpendicular to  ${\w b}\equiv {\w B}/B$,  the unit vector parallel to Earth's magnetic field ${\w B}$, and $\sigma_H$ is the Hall conductivity perpendicular to both, the electric and magnetic fields.  The expressions for the components of ${\w \sigma}$ are $\sigma_\| = e^2n_e/m_e\nu_c=\epsilon_0\omega_{pe}^2/\nu_c$ parallel to ${\w b}$, and
\begin{equation}\label{tre-eq-2}
\sigma_P=\frac{\nu_c^2}{\nu_c^2+\omega_{ce}^2}\sigma_\|, \qquad \sigma_H=-\frac{\nu_c\,\omega_{ce}}{\nu_c^2+\omega_{ce}^2}\sigma_\|,
\end{equation}
where $\omega_{ce}= eB/m_e$ and $\omega_{pe}=e\sqrt{n/ \epsilon_0m_e}$ are the electron cyclotron and plasma frequencies, respectively, and $\epsilon_0$ is the dielectric constant of free space. Below the mesosphere these expressions simplify to $\sigma_P=\sigma_\|, \sigma_H\simeq -\omega_{ce}/\nu_c\ll 1$. Hence the atmosphere is practically isotropic with zero Hall conductivity. As Figure \ref{tre-fig-temper} suggests, this changes  at altitudes above 90 km at the top of the mesosphere. Here the collision frequency drops below the electron cyclotron frequency (the vertical line in the figure) which has a weak altitude dependence only according to the weak variation of the Earth's magnetic field $B\simeq B_0(1-3h/{\rm R_E})$ with height $h$ through the atmosphere; $B_0$ is the value at the Earth's surface on a particular geomagnetic field line. The full conductivity tensor comes into play, and this has consequences for the development of electric discharges at these altitudes. The conductivity $\sigma_\|$ is low in the atmosphere meaning that the atmosphere is a poor conductor. Under discharge  conditions this may change, though. In contrast, the ions are practically collisionally isotropic because $\nu_c\gg\omega_{ci}$ by far exceeds the ion cyclotron frequency $\omega_{ci}=Z(m_e/m_i)\,\omega_{ce}$ up to the altitudes of the ionosphere ($Z $ is the ionic charge number, and $m_i$ the ion mass).

Introducing the more detailed papers collected in Chapter 7  on the most interesting special cases of atmospheric discharges in the upper atmosphere and their effects on radio wave propagation, we will, in the following, briefly review the types of electrical discharges that may be relevant under gaseous atmospheric conditions. We explicitly exclude here other important discharges such as those in \index{storms!dust} charged dust and sand storms \index{storms!sand} which are considered in detail elsewhere in this volume \cite{Renno2008}. On Earth they occur in volcanic eruptions which inject  large amounts of dust and ashes into the atmosphere where they trigger lightning. They are also known to occasionally occur in sand storms in the big terrestrial sand deserts. In the atmospheres of the dusty planets Mars and Venus they become important if not dominant \cite{Melnik1998,Farrell1999}. 

\begin{figure}%
\centerline{\includegraphics[width=8.5cm]{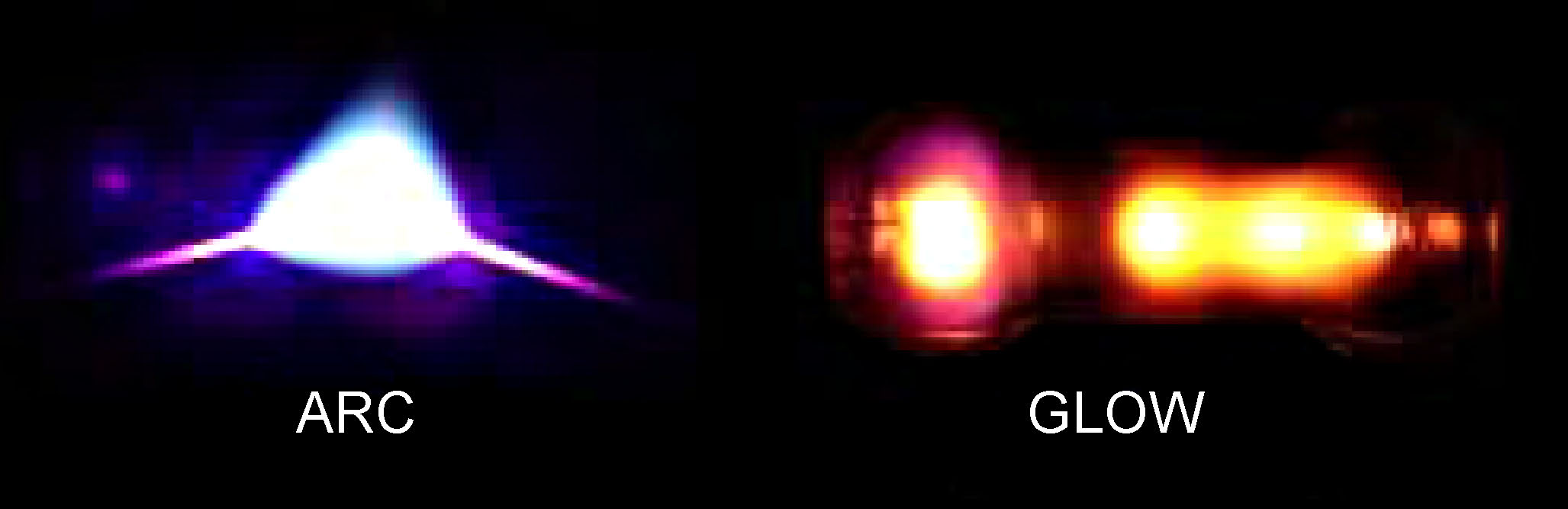}} \caption{Two types of electric discharges. {\it
Left}: A typical arc discharge, and {\it Right}: a glow discharge in an evacuated glass tube. Glow discharges are quiet and low temperature which is in contrast to the noisy and unsteady spark discharges. Because of this reason they can be used for illumination. Arc discharges, on the other hand, develop high temperatures and find technical application in welding.} \label{tre-fig-arc}
\end{figure}

\vspace{-0,2cm}
\section{Different Types of Discharges}
\noindent
In order to ignite an electric discharge in an otherwise neutral gas like the atmosphere, electrons have to be set free by some process. Subsequently an electric field must be present that can accelerate electrons to energies far above the ionisation energy fast enough in order to cause an avalanche of newly generated electrons by collisional (or other kinds of) ionisation. For this to happen the electric field must be strong, which implies that by some external force the initially present negatively and positively charged layers must become separated over a sufficiently large distance $L$. Processes capable of providing such initial charge distributions are discussed elsewhere in this issue. One of the basic processes is the weak ionisation of the atmosphere produced by the continuous inflow of highly energetic cosmic rays into the atmosphere and the cascades of nuclear fission products and elementary particles which are produced by them. Charge layer separation in a thunderstorm is a complicated process closely related to the physics of clouds and their internal dynamics. The reader is referred to the respective sections of this volume for information.
\subsection{Townsend, glow, and spark discharges} \index{discharge!spark} \index{discharge!Townsend}\index{discharge!glow}\index{discharge!corona}
Once sufficient charge has accumulated to form a charge layer and oppositely charged layers have become separated from each other, a macroscopic electric field is built up. As long as this field is weaker than some threshold field $E_{\rm crit}$, discharges\index{field!threshold} go on only slowly due to recombination and may be balanced by newly created ionisation. However, when the electric field exceeds the threshold, a violent discharge process sets in, causing breakdown and the electric field is shorted out.  In laboratory gas discharge physics one distinguishes a number of discharges. Two of them, which are accompanied by light emission, are shown in Figure \ref{tre-fig-arc}. To the left is the typical blue arc discharge between cathode and anode, and to the right is a typical red glow discharge observed in a glass tube at low pressure. 

In addition there are dark (or Townsend) discharges which, as the name implies, are not accompanied by visible light emission. Moreover, one distinguishes non-stationary discharges in connection with spark discharges. In Townsend dark discharges the current flow is rather weak and the current densities are low; the voltage remains constant over the time of the discharge. When the current strength increase the Townsend discharge makes the transition to a  glow discharge. Now the energy of the flowing electrons is large enough to excite an atom collisionally, which emits light from the infrared to the visible and up to the ultraviolet. Arc discharges acquire large current densities, strong electric fields, and may emit radiation deep into the ultraviolet and even weak X-rays. These processes have been described in textbooks \cite[for instance]{Lieberman1994,Raizer1997}.  In atmospheric physics of most interest are discharges of the kind which lead to breakdown as, for instance, when lightning discharges are generated. 

\subsection{Electron avalanches} \index{ionization!avalanche}
\noindent For the purposes of atmospheric discharges the formation of current carrying electron avalanches is  most important, as shown in Figure \ref{tre-fig-aval}. The continuous slow discharges of electric fields produced by the persistent inflow of cosmic rays into the atmosphere provide the seed population of electrons for the ignition of an avalanching discharge. In an avalanche the electron number density increases exponentially since d$n_e$/d$t=\alpha_in_e$, the production of electrons being proportional to the existing electron number density and with the proportionality factor being the Townsend ionisation rate coefficient $\alpha_i$ \cite{Raizer1997}. This relation has the solution $n_e(t)=n_e(0)\exp [\int_0^t {\rm d}t'\alpha_i(t')]$. The exponential factor indicates the temporal multiplication rate of the avalanche, which also  corresponds to the spatial multiplication rate over the distance of the evolution of the avalanche. 
\begin{figure}%
\centerline{\includegraphics[width=9cm]{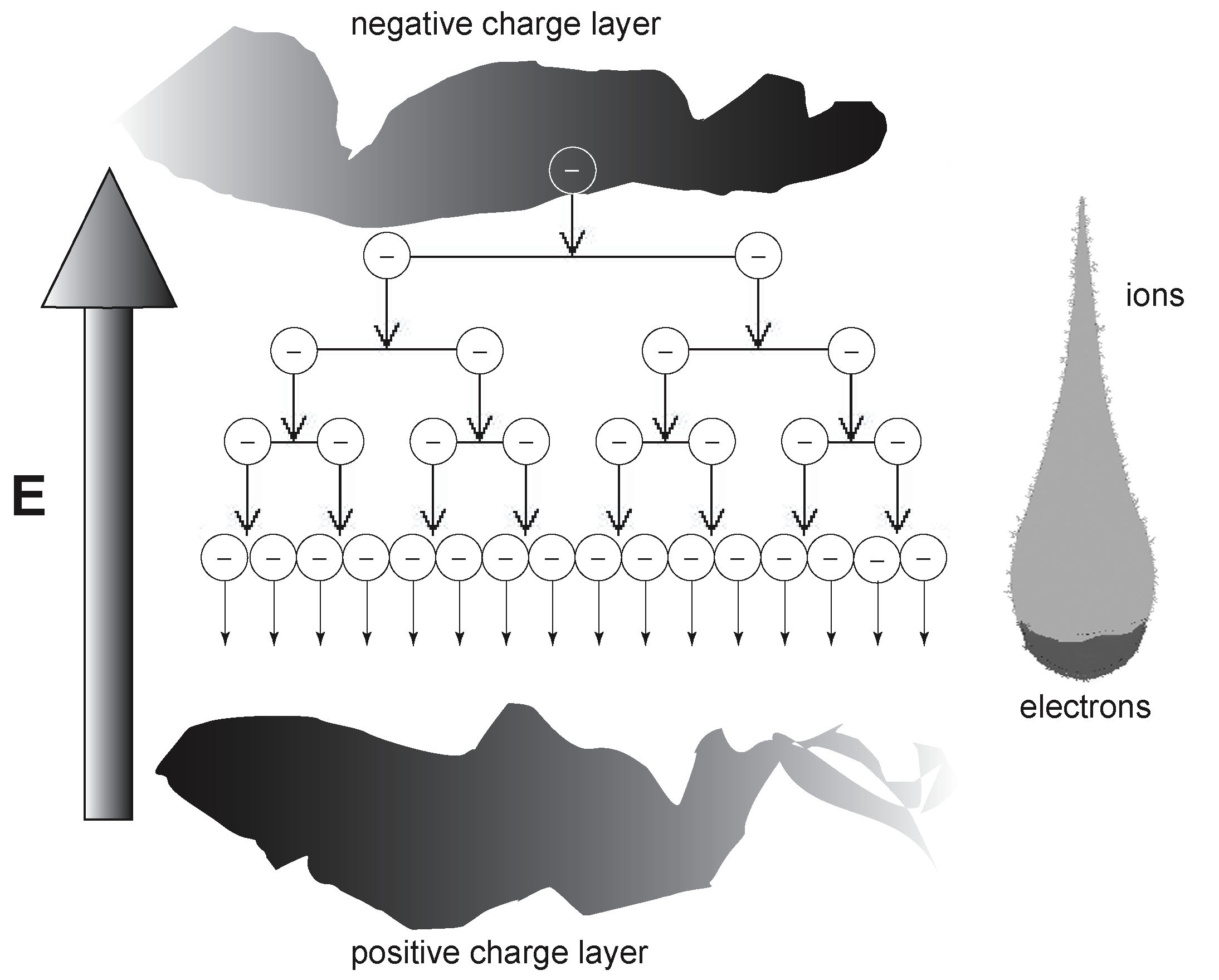}} \caption{{\it Left}: Avalanche generation in electron collisions with neutrals in presence of a sufficiently strong electric field (the Townsend process). {\it Right}: The resulting charge distribution in the avalanche. The head of the avalanche consists of fast and dense electrons. An extended wake of slow ions is left behind which try to follow the electrons but are accelerated in the opposite direction by the externally applied electric field which, however, is partially cancelled inside the ion cloud wake. }\label{tre-fig-aval}
\end{figure}

Avalanches\index{particles!avalanche} have a typical structure with a negative head consisting of the fast electrons and an extended positively charged tail containing the slow freshly produced ions which tend to retard the advancement of electrons but are kept back by the external electric field that accelerates them in the opposite direction. It is clear that in such a process the electric field must become strong enough in order to overcome recombination collisions and accelerate electrons in one mean free path length (see Figure \ref{tre-fig-temper}) up to ionisation energies. The latter are of the order of a eV, for hydrogen typically ${\cal E}_{\rm ion} \simeq 13.6$ eV. The accelerated electrons must be able to both run away and ionise. From this reasoning follows a simple condition for the necessary strength of the external field ${\w E}$ for igniting a gaseous electric discharge
\begin{equation}\label{tre-eq-3}
|{\w E}| \gtrsim E_{\rm crit} \equiv 2({\cal E}_{\rm ion}/e\lambda_{\rm mfp}).
\end{equation}
At 10 km altitude this field is quite large, amounting to $>(2-3)\times 10^6$ V\,m$^{-1}$, but decreases with altitude deep into the mesosphere ($\sim 70$ km altitude) by roughly four orders of magnitude (see Figure \ref{tre-fig-temper}) to become only a few 100 V m$^{-1}$ there. Hence, in the higher atmosphere considerably smaller electric field strengths are required in order to ignite an electric discharge than in the troposphere. The problem therefore consists less in exceeding the threshold at these altitudes than in generating charged layers with an electric field between them which requires vertical transport. But once this is achieved at lower altitudes  breakdown at higher altitudes becomes possible. In this respect it is noteworthy that sprites and jets have been observed to be closely related to low altitude thunderstorm discharges starting from the tops of thunderstorm clouds at $<$ 10 km. 

The high threshold electric field must in fact be exceeded at low altitudes of a few km, which poses a serious problem. A solution has been proposed when it was theoretically realised and proved by kinetic simulations  \cite{Gurevich1992, Roussel1994, Lehtinen1997, Babich1998} that a proper kinetic description of the avalanching process should take into account not only the generation of secondary electrons but also the production of high energy run-away electrons\index{particles!run-away electrons}. This causes what these authors call ``runaway breakdown" ({\footnotesize RBD})\index{discharge!run-away breakdown RBD}. A complete review of the theory can be found elsewhere \cite{Gurevich2001}. The idea \index{ionization!runaway breakdown} is that a small number of fast seed electrons initially reach high energy ($\gtrsim$\,10\,keV) in the electric field.  At these energies the electrons are fast enough to ignore the neutral state of matter. They take the outer shell valence electrons and the nuclei of the air molecules as free particles. For them the ordinary collisional cross section $\sigma_{en}$ for ionisation is replaced with the energy dependent Rutherford-Coulomb cross section $\sigma_{\,\rm C}$ when passing through the interior of the neutral molecules. This drastically increases the efficiency of ionisation since the Coulomb cross section is inversely proportional to the square of the particle energy; it decreases with energy which allows for energetic particles after collision to ``run away" in the field, and obey a power law distribution function $f({\cal E})\propto {\cal E}^{-1.1}$ at energies below some threshold energy above which they lose their motional energy by radiation. It has been shown \cite{Gurevich1992} that this self-consistent power law distribution is a function of the ratio of the electric to breakdown-threshold electric fields. 

The {\footnotesize RBD} threshold electric field as a function of altitude $h$ is proportional to the neutral atmospheric density 
\begin{equation}\label{tre-eq-4}
|{\w E}| \gtrsim E_{\rm crit, RBD} \simeq 0.2\, \left[\frac{n_n(h )}{n_0}\right]=\,2.0\,\exp\left[-\frac{m_ngh}{k_BT_n(h)}\right]\times 10^5\,\frac{\rm V}{\rm m}.
\end{equation}
This value is one order of magnitude less than the value obtained from Eq.\,(\ref{tre-eq-3}). Due to the dependence of $n_n(h)$ on altitude, it decreases exponentially with $h$, and at higher altitudes becomes very low indeed. The Townsend coefficient now becomes a function of the ratio of electric field to critical electric field \cite{Gurevich1992}
\begin{equation}\label{tre-eq-5}
\alpha_{i,{\rm RBD}}\simeq 0.07(E/E_{\rm crit, RBD})^\frac{3}{2},
\end{equation}
showing the pronounced nonlinearity of the self-feeding process, i.e. avalanche formation. The numerical factor 0.2  in Eq.\,(\ref{tre-eq-4}) has been determined from numerical simulations of runaway discharges at different pressures.

\vspace{-0.2cm}
\section{Secondary Effects}
\noindent The avalanches produced in these spark discharges in the atmosphere cause a number of secondary effects. Sparks are bright, emitting light due to the excitation of both atoms and ions which emit in various lines of the electromagnetic spectrum. Moreover, space charges that are built up over macroscopic vertical and horizontal length scales by the dynamics of the neutral atmosphere due to moving the charges act back on the dynamics, braking it due to the large electrostatic forces that develop when the charges become sufficiently large.  Finally, the ionisation of the atmosphere  affects the atmospheric chemistry. Other effects are the excitation of several kinds of waves, the generation of radio waves \index{waves!VLF} \index{waves!radio} and modification of propagation conditions. These have been observed to occur during and in the wake of thunderstorms and in coincidence with sprites and jets  \cite{Bosinger2008,Mika2008,Simoes2008}. Low frequency electromagnetic radiation ({\footnotesize VLF, ELF, ULF}, Schumann resonances) which accompanies {\footnotesize TLE}s is well understood \cite{Rodger1999,Reising1996,Cummer1997,Fullekrug2001,Fullekrug1998a,Fullekrug1998b,Fullekrug2006a}.  In the following we list some secondary {\it microscopic} processes in avalanches but are less familiar. Some of them might affect the physics of the discharge and might result in observable effects.

\subsection{Waves}\index{waves}
\paragraph*{Ion sound.} The simplest wave type is the mechanical distortion of the atmosphere during discharges which cause thunder and mechanical oscillations of the atmosphere. During such disturbances the ionic component of the atmosphere undergoes similar distortions which may result in low frequency ($0\!<\!\omega_{is}/\omega_{pe}\!\!<\!\!\sqrt{m_e/m_i}$) sonic waves $\omega_{is}=kc_{is}$ of speed $v_i\!\ll\! c_{is}\simeq \!\sqrt{m_e/m_i}\,v_e\!\sim \!40$\,km\,s$^{-1}$, where $v_e,v_i$ are the thermal velocities of electrons and ions, respectively. 

It is usually believed that these waves are strongly damped due to the high collision frequency at low altitudes (see Figure \ref{tre-fig-temper}). Landau damping can be neglected because of the low ion temperature. At higher altitudes than at the top of the mesopause ion sound turbulence \index{waves!ion sound} should accompany any discharge. We must compare the mean free path of electrons with the Debye length $\lambda_D\!\!=\!\!v_e/\omega_{pe}$, where the plasma frequency $\omega_{pe} \simeq 60\sqrt{n_e}$ s$^{-1}$, and the electron density is measured in m$^{-3}$. The electron density produced in a discharge is some small altitude-dependent fraction $\zeta(h)\!\ll\! 1$ of the neutral density  $n_n(h)$. The ratio of the Debye length to the mean free path $\lambda_{\rm mfp}=1/n_n(h)v_e$ is $\lambda_D/\lambda_{\rm mfp}\sim 10^{-2}\sqrt{(k_BT_e/e)(n_n(h)/n_0\zeta(h))}<1 $ for any finite electron temperature of a few 10 eV, and $\zeta(h)<n_n(h)/n_0$. Hence, the plasma properties dominate inside the avalanche. 

Moreover, the electrons being of energy $\gtrsim\!\! 10$\,eV, are much hotter than the ions, $T_e\!\gg\! T_i$. They carry the current $j\!=\!-en_eV_d$, where $V_d$ is the current drift velocity. This makes excitation of ion sound waves probable if only the growth rate $\gamma_{ia}\!>\!\nu_{c}$ exceeds the collisional electron damping rate. In an avalanche the latter is given by the Spitzer collision frequency $\nu_{\,\rm C}$. This condition, which can be written $1\!>\! V_d/v_e\!>\!200/N_D$, is always satisfied, then. With $N_D\!=\!(n_e\lambda_D^3)^{-1}\!\gg\! 1$ the number of electrons in a Debye sphere, its right-hand side is a small number. The ion-sound waves cause the avalanche to develop magnetic field-aligned striations \cite[e.g., for the discussion of the instabilities and their nonlinear evolution]{Davidson1972,Treumann1997}. 

\paragraph*{Buneman two-stream instability.} For $V_d>v_e$ the avalanche plasma is Buneman\index{Buneman, Oscar} two-stream unstable \cite{Buneman1958}. {\footnotesize RBD} discharges with their comparably high electron density and electron kinetic energy \cite{Gurevich2001} are candidates for the two-stream instability. The two-stream instability\index{instability!two-stream} is riding on the beam, i.e. it propagates at about the same velocity as the beam, being for long time in resonance with the current carrying electrons and thus giving rise to high wave intensities. This instability saturates by formation of localised electron\index{particles!electron holes} (and also ion) holes, which correspond to locally (a few 10 Debye lengths long) very strong electric fields. This has been demonstrated in numerical simulations \cite{Newman2002}. [The existence of holes has been confirmed in the auroral plasma  \cite{Carlson1998,Ergun1998,Elphic1998}.] Depending on its polarity, the localised electric field traps electrons or ions. The trapped avalanche electrons bounce back and forth in the holes and become violently heated, dissipating a substantial part of the beam current and electric field energy. Conversely, passing (untrapped) electrons are accelerated further and are at the same time cooled to low temperatures, as seen in Figure\,\ref{tre-fig-newm}. 
\begin{figure}
\centerline{\includegraphics[width=8cm]{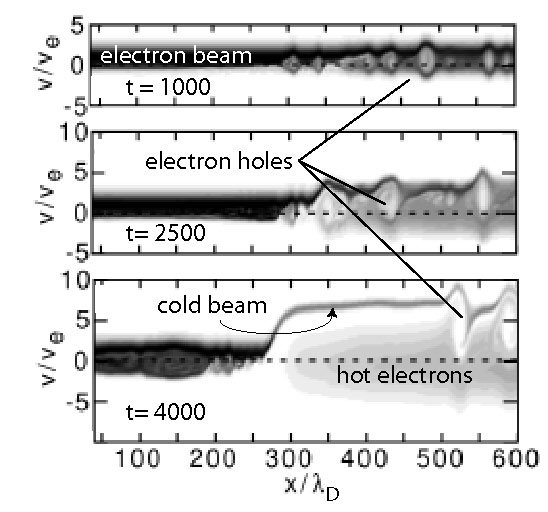}}
\caption{Electron phase space plots at three successive simulation times (measured in plasma times $\omega_{pe}^{-1}$) of the formation of electron holes in a two-stream unstable plasma. Instability is forced by the presence of a density dip (potential wall for electrons) in the centre of the simulation box which interacts with an injected electron beam current (after Newman et al. 2002). The box has one space (measured in Debye length) and one velocity coordinate (measured in electron thermal speeds $v_e$. Hole formation is seen on the right. The holes move to the right with beam velocity. They trap electrons and cause strong electron heating of the trapped electrons. In addition acceleration of a narrow (cold) beam is observed which escapes from the hole region at about 7 times the initial velocity.}\label{tre-fig-newm}
\end{figure}

\paragraph*{Phase space hole effects.} The holes resulting from the two-stream instability are regions of high electric field pressure which affects the equilibrium between the avalanche and its neutral environment. Force balance requires $\nabla_nP_n\!=\!-\nabla_eP_e\!-\nabla_EP_E$, where $P_n\!=\!n_nk_BT_n$ is the atmospheric pressure, $P_e\!=\!n_ek_BT_e$ is the electron pressure in the avalanche, and $P_E\!=\!\frac{1}{2}\epsilon_0|{\bf E}|^2\{\partial[\omega\epsilon(\omega)]/\partial\omega\}$ is the field pressure. The electron and field pressures together in the avalanche act to expel the neutral gas from the avalanche region. But the forces on the right and left act on different scales. The scale height $L_n$ of the pressure is several kilometers, while the scales $L_e\!\sim\! L_E$ of the holes are only a few 10\,Debye lengths long. The local force balance thus becomes
\begin{displaymath}
P_n=\frac{L_n}{10\lambda_D}(P_e+P_E).
\end{displaymath}
Moreover, $P_n\!\simeq\! 3\times\! 10^4\! n_n(h)/n_0$\,J\,m$^{-3}$. Assume that the electrons have an energy of\,10 keV (just a fraction of the energy assumed in Gurevich's {\footnotesize RBD} mechanism). Then, $P_e\!\simeq\! 10^{-15}n_e$\,J\,m$^{-3}$. To calculate the electric field pressure one needs the wave dielectric constant $\epsilon(\omega)$. For the two-stream instability we have $\partial[\omega\epsilon(\omega)]/\partial\omega\!\sim\! \omega_{pe}^2/(0.03\,\omega_{pe})^2\!\sim\! 10^3$. Thus, $P_E\!\simeq\! 5\!\times\! 10^{-9}|E|^2$\,J\,m$^{-3}$. The Debye length is $\lambda_D\!\sim\! (0.7/\!\sqrt{n_e})$\,m. The balance equation becomes
\begin{displaymath}
10^4n_n(h)\sim 10^2 L_n({\rm km})n_e^\frac{1}{2}(10^{-15} n_e+5\times 10^{-9}|E|^2).
\end{displaymath}
This condition can be satisfied with moderate avalanche densities $n_e\!\!<\!10^{10}\,{\rm m}^{-3}$ and electric fields $|E|\!\sim$\,a few\,V\,m$^{-1}$. Stronger fields and higher avalanche densities will completely blow the neutral atmosphere away from the region of the hole. This crude estimate shows that the localisation of the electric field in electron holes and trapping of electrons as a consequence of the two-stream instability causes a substantial local dilution of the neutral atmospheric gas in the holes in the avalanche. It causes a filamentation of the avalanche into striations and narrow packets of trapped electrons and strong fields. 

Fine structuring has been observed in streamers \cite{Ebert2006} though this important observation is interpreted differently. Ebert et al. (2006) describe their observations as a multiscale structure of streamers occurring mainly in the head of the streamers. We should, however, note that the holes ride on the beam and move with about beam velocity. Thus the effect described here also happens mainly in the narrow head layer of the avalanche where the electron density, energy and velocity are highest.

\paragraph*{Local whistlers.} \index{waves!whistler} 
Avalanches can serve as sources of locally excited whistlers. This is obvious from considering the trapped electron distribution in Figure\,\ref{tre-fig-newm}. The electrons that are trapped in the electron hole potentials bounce back and forth along the magnetic field. Due to their gyrational motion\index{particles!gyration} in the geomagnetic field they possess magnetic moments $\mu_e\!=\!T_{e\perp}/B$ ($T_{e\perp}$ is the electron temperature perpendicular to the geomagnetic field). As a consequence of the bounce motion\index{particles!bouncing} of the trapped electrons, the electron holes\index{waves!electron holes} become unstable \cite{Ergun1998} with respect to whistlers of frequency $\omega_{ci}\!\ll\!\omega\!\ll\!\omega_{ce}$. While moving with the parallel hole velocity along the magnetic field, they radiate whistlers acting like point sources. In a spatially varying magnetic field this parallel motion gives rise to a  saucer like frequency-time appearance of the whistlers. [Such whistler-saucer emissions were indeed observed first under the conditions in the auroral ionosphere \cite{Ergun1998}. Similar phenomena during thunderstorms have been reported  \cite{Parrot2008} from the low-altitude spacecraft {\footnotesize D}{\scriptsize EMETER}.\index{satellites!DEMETER}] Their propagation direction is oblique to the geomagnetic field at the whistler-resonance cone,  \index{waves!saucers} with frequency near the lower-hybrid frequency $\omega_{lh}\!\simeq\! \omega_{pi}/(1\!+\!\omega_{pe}^2/\omega_{ce}^2)^\frac{1}{2}$. In the strong geomagnetic field this is close to the ion plasma frequency $\omega_{pi}$. Their frequency-time dependence maps the avalanche plasma density along the whistler path. Propagating obliquely, these short-perpendicular long-parallel wavelength whistlers are trapped in the avalanche. Being multiply reflected from its boundaries, the whistlers ultimately escape on the top-side with Alfv\'en velocity $V_{A}\!=\!B/\!\sqrt{\mu_0m n_e}$ mostly in perpendicular direction to the magnetic field thereby spreading over a large hollow horizontal area of narrow opening angle, centred around the source before being absorbed.  Excitation of the mesospheric or exospheric gas might lead to emissions over the whole area of these whistlers resembling the Elve emission. 

\subsection{Radiation}\index{radiation}

\noindent Several types of electromagnetic radiation\index{radiation!electromagnetic} have been observed in relation to atmospheric electric discharges, both natural and artificially triggered \cite{Dwyer2003} lightning, in addition to the above mentioned spherics and the whistlers propagating in the ionosphere and magnetosphere\index{magnetosphere} caused by them. Lightning discharges have been observed to be accompanied by radiation in the optical,  {\footnotesize UV, X} rays and even  up to the gamma ray energy range and, also in the {\footnotesize ELF} and {\footnotesize ULF} radio wave ranges \cite{Simoes2008,Bosinger2008,Mika2008}. Moreover, spacecraft observations of other planets (Jupiter, Saturn) provide evidence of short duration radio emissions in connection with lightning on those planets \cite{Fischer2008}. \index{radiation!X-rays} \index{radiation!gamma-rays} \index{waves!Schumann resonance}

\paragraph*{Gamma- and X-rays\ {\rm(}Bremsstrahlung{\rm )}.} Referring to the above discussion of the general physics of breakdowns it can be expected that radiation is emitted from breakdown discharges in the optical to {\footnotesize X}-ray ranges. An early prediction \cite[1956]{Wilson1925} based on the Townsend mechanism that lightning should be accompanied by the generation of energetic electrons and possibly by the emission of {\footnotesize X}-rays has been confirmed by the more recent \index{BATSE} {\footnotesize B}{\scriptsize ATSE} experiment aboard the \index{satellites!Compton GRO} Compton Gamma Ray Observatory \cite{Fishman1994} and by the \index{satellites!Alexis} Alexis spacecraft observations  \cite{Blakeslee1989,Holden1995}. Gamma radiation is believed to be caused by the energetic electron component generated in the runaway phase of the discharge when avalanches of electrons are formed. In the light of the above discussion on holes it is the fast cold high energy electron beam that is responsible for their emission. The radiation mechanism has been investigated theoretically \cite{Gurevich1992,Roussel1994}.  Since very high energy electrons are required to generate these Gamma ray bursts the observations completely falsify Wilson's mechanism while they are in excellent agreement with the {\footnotesize RBD} mechanism, providing a strong argument in favour of runaway breakdown and lowering of the avalanching threshold. The gamma rays observed are not caused in a nuclear interaction, and no gamma lines have been detected yet. They form the high energy tail of the energetic X ray emission resulting from free-free (`bremsstrahlung') radiation of the most energetic runaway electrons. The absence of lines is an indication for the existence of an upper threshold of the accelerating electric field. 
\begin{figure}
\centerline{\includegraphics[width=12cm]{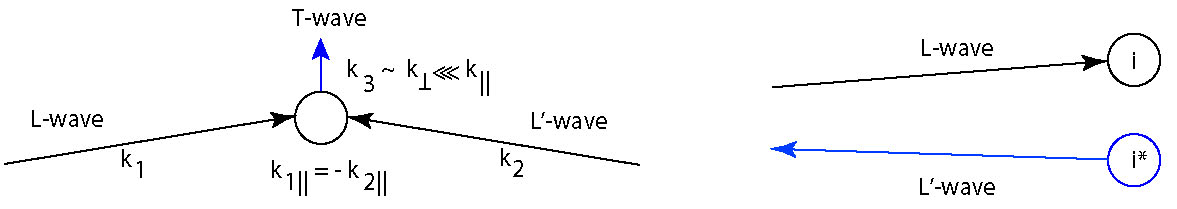}}
\caption{{\it Left}: The {\sf L\,+\,L}$^\prime \to$ {\sf T} plasma wave process for generation of electromagnetic radiation {\sf T}. The large parallel wave numbers $k_\|$ of the counterstreaming Langmuir waves cancel, and a transverse wave of frequency $\omega_3\!=\!\omega_1\!+\!\omega_2\!\approx\! 2\,\omega_{pe}$ (with very small perpendicular wave number $k_3\!\sim\! k_\perp\!\lll\! k_\|$ and therefore large phase and group velocities $\sim\! c$) is emitted. \  {\it Right}: Scattering of an {\sf L}-wave by an ion {\sf i} as a process in which an {\sf L}$^\prime$-wave is generated. {\sf i}$^*$ is the excited ion which is left over after the `collision' with the {\sf L}-wave.} \label{tre-fig-ll}
\end{figure}

\paragraph*{Radio waves: Plasma processes.} Any emitted high-frequency radio radiation can hardly be via the synchrotron process \index{radiation!sychrotron}  \cite{Rybicki1979} as this requires much higher emission measures, i.e. large volumes  occupied by energetic electrons in addition to much stronger magnetic fields, than those which are available. The only imaginable mechanisms are  plasma mechanisms like the head-on interaction of beam generated Langmuir ({\sf L}) waves \index{waves!Langmuir} due to a process {\footnotesize {\sf L\,+\,L}$^\prime  \to$\,{\sf T}} (where {\footnotesize {\sf T}} is the transverse long-wavelength radio wave). This process is shown schematically in Figure \ref{tre-fig-ll}.

Generated by the cold fast  electron beam, {\footnotesize {\sf L}}-waves propagate in the direction of the beam. The backward moving {\footnotesize {\sf L}$^\prime$}-wave can be generated by backscattering of the {\footnotesize{\sf L}}-wave by the slow cold ion component according to the process {\footnotesize{\sf L+i}$\to${\sf L$'$+i}}$^*$. Here {\footnotesize{\sf i}} is the  involved ion, {\footnotesize{\sf L}$'$} the backscattered Langmuir wave, the wave vector direction of which is reversed, and {\footnotesize{\sf i}}$^*$ is the excited ion (which emits light).  {\footnotesize{\sf L}}-waves can also interact with low frequency ion sound waves ({\footnotesize{\sf IS}}) according to the reaction {\footnotesize{\sf L+IS} $\to${\sf T}}, following the same scheme as shown in Figure \ref{tre-fig-ll}. This process has a lower efficiency than the {\footnotesize{\sf L+L}$'$} process, though. In the former case the frequency of the emitted radio wave is close to twice the plasma frequency $\omega_3({\footnotesize\textsf{L+L}})\!\simeq\! 2\omega_{pe}$. Because the frequency $\omega_{is}\!\ll\!\omega_{pe}$ of the {\footnotesize{\sf IS}}-wave is much smaller than the electron plasma frequency, the process {\footnotesize{\sf L+IS}} emits  radiation of frequency near the plasma frequency: $\omega_3({\footnotesize\textsf{L+IS}})\!\sim\!\omega_{pe}$. For avalanche beams of density $n_e\sim 10^{10}$\,m$^{-3}$ this frequency is about $\sim$1\,MHz.  \index{waves!wave-wave interaction}\index{waves!decay}\index{instability!decay}\index{waves!ion sound}\index{waves!ion scattering}

Excitation of electron-acoustic waves ({\footnotesize{\sf EA}}) of frequency close to a fraction of $\omega_{pe}$ can also contribute to HF-radio emission. The reaction equation is {\footnotesize{\sf L+EA}$\to${\sf T}}, and the frequency is above but close to $\omega_{pe}$, typically $\omega_3({\footnotesize{\textsf{L+EA}}})\!\!\lesssim\!1.5\,\omega_{pe}$.  All these mechanisms depend crucially on the presence of a hot plasma electron component, of cool ions, and only a small admixture of cold background electrons. These requirements imply that a substantial fraction of the neutral atmospheric components has been locally expelled from the avalanche region. \index{radiation!plasma}\index{instability!maser}

\paragraph*{Radio waves: Electron-cyclotron maser.} Very intense HF-radio waves can be generated by the electron-cyclotron maser \cite[for a recent review]{Treumann2006}. \index{radiation!maser}  This is an extraordinarily efficient process. Its requirements are a strong field-aligned electric potential difference, parallel current flow and a strong geomagnetic field in addition to a negligible component of cold electrons and neutrals. It also requires some pitch angle scattering of the electron distribution such that the electrons have a substantial velocity component perpendicular to the geomagnetic field. Again, this mechanism can be realised only if the neutral atmospheric gas is expelled from the avalanching electron cloud. The electron-cyclotron maser is a direct process in which no other catalysing waves are involved. It generates long wavelength ($\lambda_{\rm ecm}\!\! \sim$\,few\,100 m) electromagnetic radiation at the local electron cyclotron frequency $\omega_{\rm ecm}\!\sim\!\omega_{ce}$, which in the atmosphere is in the MHz range. The bandwidth of the emitted radiation depends on the electron beam temperature ranging from very narrow ($\ll\! \omega_{ce}$) up to a substantial part of the electron cyclotron frequency.


}
\end{article}

\printindex


\vspace{-0.3cm}
\begin{thebibliography}{}
{\scriptsize
\bibitem[\protect\citeauthoryear{Aguilar-Benitez et al.}{1990}]{Aguilar1990}
M. Aguilar-Benitez et al., Reviews of Particle Properties, Phys. Lett. B 239 (April 1990)

\bibitem[\protect\citeauthoryear{Appleton}{1947}]{Appleton1947}
E. V. Appleton, in Nobel Lectures 1947, Physics 1942-1962 (Elsevier, Amsterdam, 1964) pp. 79-86

\bibitem[\protect\citeauthoryear{Babich et al.}{1998}]{Babich1998}
L. P. Babich et al., 
Phys. Lett. A 245, 460 (1998)

\bibitem[\protect\citeauthoryear{Balmer}{1885}]{Balmer1885}
 J. J. Balmer, Annalen Phys. Chem. 25, 80-85 (1885)
 
\bibitem[\protect\citeauthoryear{Baumjohann and Treumann}{1996}]{Baumjohann1996}
W. Baumjohann, R. A. Treumann, Basic Space Plasma Physics (Imperial College Press, London 1996)

\bibitem[\protect\citeauthoryear{Bernshtam et al.}{2000}]{Bernshtam2000}
V. A. Bernshtam, Yu. V. Ralchenko, Y. Maron, J. Phys. B: Atm. Mol. Opt. Phys. 33, 5025-5032 (2000)


\bibitem[\protect\citeauthoryear{Bosinger and Shalimov}{2008}]{Bosinger2008}
T. Bosinger, S. Shalimov,  \ssr\ this issue (2008)

\bibitem[\protect\citeauthoryear{Blakeslee et al.}{1989}]{Blakeslee1989}
R. J. Blakeslee, H. J. Christian, B.  Vonnegut,  \jgr\ 94, 13135 (1989)

\bibitem[\protect\citeauthoryear{Bowers}{1991}]{Bowers1991}
B. Bowers,  A History of Electric Light and Power
(Peter Peregrinus Ltd., London 1991); --, Lightening the Day. A History of Lighting Technology
(Oxford Univ. Press, Oxford 1998)


\bibitem[\protect\citeauthoryear{Buneman}{1958}]{Buneman1958}
O. Buneman, 
\prl\  1, 8-9 (1958)


\bibitem[\protect\citeauthoryear{Carlson et al.}{1998}]{Carlson1998}
C. W. Carlson et al., \grl\ 25, 2017 (1998)

\bibitem[\protect\citeauthoryear{Chapman}{1931}]{Chapman1931}
S.  Chapman,  Proc. Phys. Soc. 43, 26 and 483 (1931)

\bibitem[\protect\citeauthoryear{Compton and Langmuir}{1930}]{Compton1930}
K. T. Compton,  I. Langmuir, \rmp\  2, 123-242 (1930);  I. Langmuir, K. T.  Compton, \rmp\ 3, 191-258 (1931)


\bibitem[\protect\citeauthoryear{Crookes}{1879}]{Crookes1879}
W.  Crookes, Phil. Trans. Roy. Soc. London 170, 135 and 641 (1879)

\bibitem[\protect\citeauthoryear{Cummer and Inan}{1997}]{Cummer1997}
S. A. Cummer and U. S. Inan, Geophys. Res. Lett. 24, 1731-1734 (1997)

\bibitem[\protect\citeauthoryear{Davy}{1807}]{Davy1807}
H. Davy, Sir, Phil. Trans. Roy. Soc. London 97, 1-56 (1807)

\bibitem[\protect\citeauthoryear{Davidson}{1972}]{Davidson1972}
R. C. Davidson, Methods in Nonlinear Plasma Theory (Academic Press, New York 1972)

\bibitem[\protect\citeauthoryear{Debye and H\"uckel}{1923}]{Debye1923}
P. Debye,  E.  H\"uckel, Physikalische Zeitschrift 24, 185-206 (1923)

\bibitem[\protect\citeauthoryear{Dwyer et al.}{2003}]{Dwyer2003}
J. R. Dwyer et al., \sci\ 299, 694-697 (2003)

\bibitem[\protect\citeauthoryear{Ebert et al.}{2006}]{Ebert2006}
U. Ebert et al., Plasma Sourc. Sci. Technol. 15, S118-S129 (2006)

\bibitem[\protect\citeauthoryear{Elphic et al.}{1998}]{Elphic1998}
R. C. Elphic et al., \grl\ 25, 2025 and 2037 (1998)

\bibitem[\protect\citeauthoryear{Ergun et al.}{1998}]{Ergun1998}
R. E. Ergun et al., \grl\ 25, 2033 (1998); 28, 3805 (2001);  \pop\ 10, 454 (2003); D. L. Newman, M. V. Goldman, R. E.  Ergun, \pop\ 9, 2337 (2002)




\bibitem[\protect\citeauthoryear{Faraday}{1833}]{Faraday1833}
M.  Faraday, Experimental Researches, Series III (1833);  Phil. Trans. Roy. Soc. 124, 77 (1834)


\bibitem[\protect\citeauthoryear{Farrell et al.}{1999}]{Farrell1999}
W. M. Farrell et al., \jgr\ E4 104, 3795 (1999)

\bibitem[\protect\citeauthoryear{Fischer et al.}{2008}]{Fischer2008}
G. Fischer et al.,  \ssr\  this issue (2008)

\bibitem[\protect\citeauthoryear{Fishman et al.}{1994}]{Fishman1994}
G. J. Fishman et al., \sci\ 264, 1313 (1994)

\bibitem[\protect\citeauthoryear{Franck and Hertz}{1914}]{Franck1914}
J. Franck,  G.  Hertz, Verh. Dtsch. Phys. Ges. 16, 457-467 (1895) 

\bibitem[\protect\citeauthoryear{Franz et al.}{1990}]{Franz1990}
R. C. Franz, R. J. Nemzek, J. R.  Winckler, \sci\ 249, 48-50 (1990)

\bibitem[\protect\citeauthoryear{F\"ullekrug et al.}{1998}]{Fullekrug1998a}
M. F\"ullekrug, A. C. Fraser-Smith and S. C. Reising, Geophys. Res. Lett. 25, 3497-3500 (1998)

\bibitem[\protect\citeauthoryear{F\"ullekrug and Reising}{1998}]{Fullekrug1998b}
M. F\"ullekrug and S. C. Reising, Geophys. Res. Lett. 25, 4145-4148 (1998)

\bibitem[\protect\citeauthoryear{F\"ullekrug et al.}{2001}]{Fullekrug2001}
M. F\"ullekrug, D. R. Moudry, G. Dawes and D. D. Sentman, J. Geophys. Res. 106, 20189-20194 (2001)

\bibitem[\protect\citeauthoryear{F\"ullekrug et al.}{2006a}]{Fullekrug2006a}
M. F\"ullekrug, M. Ignaccolo and A. Kuvshinov, Radio Sci. 41, RS2S19, doi:10.1029/2006RS003472 (2006a)


\bibitem[\protect\citeauthoryear{Fullekrug et al.}{2006b}]{Fullekrug2006}
M. Fullekrug, E. A. Mareev, M. J. Rycroft (eds.) Sprites, Elves and Intense Lightning Discharges, NATO Sci. Ser. II,  225 (Springer, Heidelberg,  2006b)

\bibitem[\protect\citeauthoryear{Gillmor}{1971}]{Gillmor1971}
C. S.  Gillmor, Charles Augustin Coulomb 
 (Princeton Univ. Press, Princeton N.J., 1971)

\bibitem[\protect\citeauthoryear{Gurevich et al.}{1992}]{Gurevich1992}
A. V. Gurevich,  G. M. Milikh, R.  Roussel-Dupr{\'e}, Phys Lett. A 165, 463 (1992); A. V. Gurevich et al., A 275, 101 (2000); A 282, 180 (2001);  A 301, 320 (2002)

\bibitem[\protect\citeauthoryear{Gurevich and Zybin}{2001}]{Gurevich2001}
A. V. Gurevich, K. P. Zybin, Phys. Usp. 44, 1119-1140 (2001) 




\bibitem[\protect\citeauthoryear{Hittorff}{1879}]{Hittorff1879}
W.  Hittorff, Wiedemannsche Annalen 7, 613 (1879)

\bibitem[\protect\citeauthoryear{Holden et al.}{1995}]{Holden1995}
D. N. Holden, C. P. Munson, J. C.  Devenport, \grl\ 22, 889 (1995)

\bibitem[\protect\citeauthoryear{Kallmann}{1953}]{Kallmann1953}
H. K.  Kallmann,  Phys. Rev. 90, 153 (1953)

\bibitem[\protect\citeauthoryear{Knight}{1992}]{Knight1992}
D.  Knight, Humphry Davy, Science and Power (Cambridge Univ. Press, Cambridge UK, 1992)


\bibitem[\protect\citeauthoryear{Langmuir}{1923}]{Langmuir1923}
 I. Langmuir, Phys. Rev. 22, 357 (1923)

\bibitem[\protect\citeauthoryear{Lehtinen et al.}{1997}]{Lehtinen1997}
N. G. Lehtinen et al., 
\grl\ 24, 2639 (1997)

\bibitem[\protect\citeauthoryear{Lieberman and Lichtenberg}{1994}]{Lieberman1994}
M. A. Lieberman, A. J.  Lichtenberg, Principles of Plasma Discharges and Materials Processing (J. Wiley \& Sons Inc., New York 1994)

\bibitem[\protect\citeauthoryear{Lomonossov}{1748}]{Lomo1748}
M. V.  Lomonossov, Letter to Euler, July 5 (1748)

\bibitem[\protect\citeauthoryear{Lyman}{1914}]{Lyman1914}
T.  Lyman, \nat\ 93, 241 (1914)

\bibitem[\protect\citeauthoryear{Melnik and Parrot}{1998}]{Melnik1998}
O. Melnik, M.  Parrot, \jgr\ 103, 29107-29118,
doi: 10.1029/98JA01954 (1998)

\bibitem[\protect\citeauthoryear{Mika and Haldoupis}{2008}]{Mika2008}
A. Mika, C. Haldoupis,   \ssr\ this issue (2008)

\bibitem[\protect\citeauthoryear{Mishin and Milikh}{2008}]{Mishin2008}
E. Mishin, G. Milikh,  \ssr\ this issue (2008)

\bibitem[\protect\citeauthoryear{Mott-Smith}{1971}]{Mott-Smith1971}
H. M.  Mott-Smith, \nat\ 233, 219 (1971)

\bibitem[\protect\citeauthoryear{Nahin}{1990}]{Nahin1990}
P. J.  Nahin, Scientific Am., June issue, 122-129 (1990)

\bibitem[\protect\citeauthoryear{Neubert}{2003}]{Neubert2003}
T. Neubert, \sci\ 300, 747-749 (2003)

\bibitem[\protect\citeauthoryear{Newman et al.}{2002}]{Newman2002}
D. L. Newman, M. V. Goldman, R. E.  Ergun, \pop\ 9, 2337 (2002)

\bibitem[\protect\citeauthoryear{Parrot et al.}{2008}]{Parrot2008}
M. Parrot et al., \ssr\ this issue (2008)

\bibitem[\protect\citeauthoryear{Paschen}{1908}]{Paschen1908}
F.  Paschen, Ann. Phys. 27, 537-570 (1908)

\bibitem[\protect\citeauthoryear{Pasko}{2007}]{Pasko2007}
V. P.  Pasko, Plasma Sources Sci. Technol. 16, S13-S29 (2007)

\bibitem[\protect\citeauthoryear{Raizer}{1997}]{Raizer1997}
Yu. P.  Raizer, Gas Discharge Physics (Springer Verlag, New York, 1997)

\bibitem[\protect\citeauthoryear{Reising et al.}{1996}]{Reising1996}
S. C. Reising, U. S. Inan and T. F. Bell, Geophys. Res. Lett. 23, 3639-3642 (1996)

\bibitem[\protect\citeauthoryear{Renno and Kok}{2008}]{Renno2008}
N. O. Renno, J.  Kok, \ssr\ this issue (2008)

\bibitem[\protect\citeauthoryear{Richardson}{1908}]{Richardson1908}
 O. W.  Richardson, Phil. Mag. 16, 740 (1908); Nobel Lectures 1928 (Elsevier, 1964) pp. 224-236


\bibitem[\protect\citeauthoryear{Ritz}{1908}]{Ritz1908}
W.  Ritz, Ann. Phys. 25, 660-696 (1908)

\bibitem[\protect\citeauthoryear{Rodger}{1999}]{Rodger1999}
C. J. Rodger, Rev. Geophys. 37, 317-336 (1999)

\bibitem[\protect\citeauthoryear{Roentgen}{1895}]{Roentgen1895}
W. C.  Roentgen,  Sitzber. Physik. Med. Ges.  137, 1 (1895)

\bibitem[\protect\citeauthoryear{Roussel-Dupr{\' e} et al.}{1994}]{Roussel1994}
R. Roussel-Dupr{\'e}  et al.,  Phys Rev. E  49, 2257 (1994);  \jgr\  101, 2297 (1996)


\bibitem[\protect\citeauthoryear{Roussel-Dupr\'e et al.}{2008}]{Roussel2008}
R. Roussel-Dupr\'e et al., 
\ssr\ this issue (2008)

\bibitem[\protect\citeauthoryear{Rybicki and Lightman}{1979}]{Rybicki1979}
G. B. Rybicki, A. P. Lightman, Radiative Processes in Astrophysics (J. Wiley $\&$ Sons Inc., New York, 1979)

\bibitem[\protect\citeauthoryear{Rycroft et al.}{2008}]{Rycroft2008}
M. A. Rycroft et al., \ssr\ this issue (2008)

\bibitem[\protect\citeauthoryear{Sentman et al.}{1995}]{Sentman1995}
D. D. Sentman et al., 
\grl\  22, 1205-1208 (1995) 

\bibitem[\protect\citeauthoryear{Sim$\tilde{\rm o}$es et al.}{2008}]{Simoes2008}
F. Sim$\tilde{\rm o}$es et al., 
\ssr\ this issue (2008)

\bibitem[\protect\citeauthoryear{Thomson}{1897}]{Thomson1897}
J. J.  Thomson, Phil. Mag.  44, 293 (1897)

\bibitem[\protect\citeauthoryear{Townsend}{1901}]{Townsend1901}
E.  Townsend,  Phil. Mag.  1, 198 (1901); Electricity in Gases (Oxford Univ. Press., Oxford 1915)


\bibitem[\protect\citeauthoryear{Treumann}{2006}]{Treumann2006}
R. A.  Treumann,  Rev. Astron. Astrophys.  13, 229-315 (2006)

\bibitem[\protect\citeauthoryear{Treumann and Baumjohann}{1997}]{Treumann1997}
R. A. Treumann, W. Baumjohann, Advanced Space Plasma Physics (Imperial College Press, London 1997)

\bibitem[\protect\citeauthoryear{Watson-Watt}{1950}]{Watson-Watt1950}
R. Watson-Watt, Sir, The Scientific Monthly, December issue, 353-358 (1950)

\bibitem[\protect\citeauthoryear{Wilson}{1925}]{Wilson1925}
C. T. R.  Wilson, Proc. Royal Soc. London  37 , 32D (1925);   236 , 297 (1956)


\bibitem[\protect\citeauthoryear{Yair}{2008}]{Yair2008}
Y. Yair et al.,  \ssr\ this issue (2008)
}
\end{thebibliography}
\end{document}